\documentclass[a4paper, preprint, 12pt]{revtex4}
\usepackage{graphicx}
\usepackage{amssymb}

\newcommand{\eqref}[1]{(\ref{#1})}

\newcommand{\IEEEas}{\it IEEE Trans. on Appl. Supercond. \rm}
\newcommand{\IEEEmag}{\it IEEE Trans. on Magn. \rm}
\newcommand{\SUST}{\it Supercond. Sci. Technol. \rm}
\newcommand{\FED}{\it Fusion Eng. Des. \rm}
\newcommand{\CRYO}{\it Cryogenics \rm}
\newcommand{\APL}{\it Appl. Phys. Lett. \rm}

\begin{document}

%% Title, authors and addresses

\title{The superconducting proposal for the CS magnet system of FAST: 
a preliminary analysis of the heat load due to AC losses}

\author{N. Pompeo\dag \footnote{E-mail: pompeo@fis.uniroma3.it.}, L. Muzzi\ddag}

\affiliation{\dag Dipartimento di Fisica ''E. Amaldi'' and Unit\`a CNISM,
Universit\`a ``Roma Tre'', Via della Vasca Navale 84, I-00146 Roma,
Italy}

\affiliation{\ddag Association EURATOM-ENEA, C.R. Frascati, Via E. Fermi, 45, IT-00044 Frascati (RM), Italy}

\begin{abstract}
FAST (Fusion Advanced Studies Torus), the Italian proposal of a Satellite Facility to ITER, is a compact tokamak (R$_0$ = 1.82 m, a = 0.64 m, triangularity $\delta$ = 0.4) able to investigate non-linear dynamics effects of $\alpha$-particle behavior in burning plasmas and to test technical solutions for the first wall/divertor directly relevant for ITER and DEMO. Currently, ENEA is investigating the feasibility of a superconducting solution for the magnet system. This paper focuses on the analysis of the CS (Central Solenoid) magnet thermal behavior. In particular, considering a superconducting solution for the CS which uses the room available in the resistive design and referring to one of the most severe scenario envisaged for FAST, the heat load of the CS winding pack due to AC losses is preliminarily evaluated. The results provide a tentative baseline for the definition of the strand requirements and conductor design, that can be accepted in order to fulfil the design requirements.\\


\end{abstract}
\maketitle

\section{Introduction}

FAST (Fusion Advanced Studies Torus) is the Italian proposal for a new European satellite tokamak reactor aimed at supporting ITER activities and anticipating some DEMO relevant physics and technology issues \cite{Crisanti, Cucchiaro}. It has been conceived as a compact (R$_0$ = 1.82 m) machine working at high field (BT up to 8.5 T) and high plasma current (I$p$ up to 8 MA). Currently, FAST magnetic system is designed with 18 TF (Toroidal Field), 6 CS (Central Solenoid) and 6 PF (Poloidal Field) resistive coils, cooled by helium gas flow at 30 K. A feasibility study to verify whether a superconducting (SC) solution for the whole magnetic system would be possible or not, avoiding any major modifications to the machine geometry or scenarios, is currently under study at ENEA \cite{DiZenobio}. This paper focuses on the proposed superconducting solution for the CS magnet in order to assess its feasibility. Since high operative currents together with both high fields and high field rates are envisaged during the operation of the CS, the evaluation of the AC losses heat loads and the corresponding thermal behavior of the SC cables are mandatory. Therefore, starting from the layout proposed in \cite{DiZenobio}, based on the 7.5 T H-mode reference scenario of FAST \cite{Crisanti, Cucchiaro}, the heat load of the CS superconducting cable originated by AC losses is evaluated in order to provide both a tentative baseline for the definition of the strand and cable requirements and the inputs needed for a future, complete thermohydraulic study.

\section{Definition of the problem}
\subsection{Design details}

\begin{figure}[hbt!]
% Requires \usepackage{graphicx}
\centerline{\includegraphics[width=10cm]{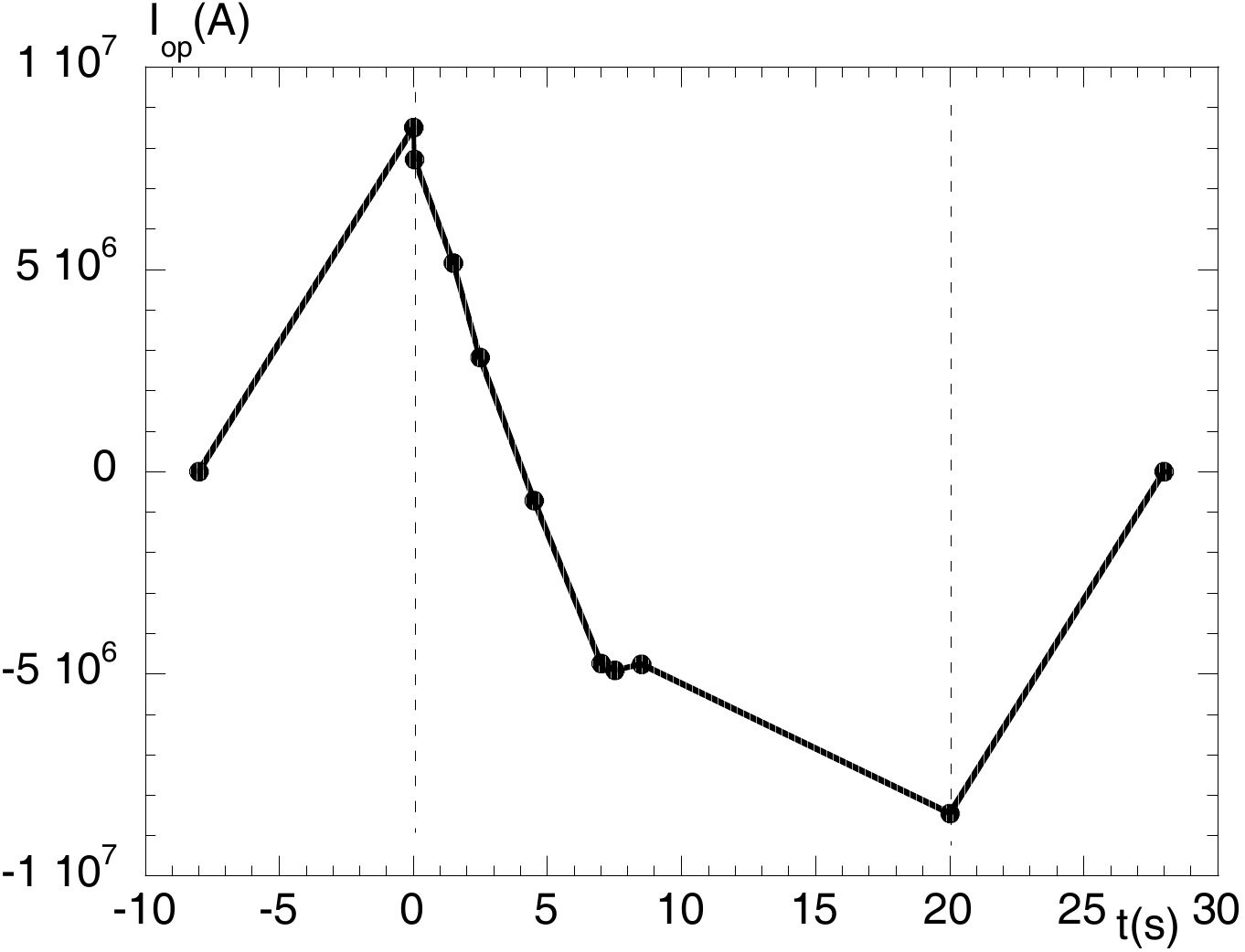}}
  \caption{Operative current vs time, in CS1 modules}
\label{fig:correnti}
\end{figure}

The SC feasibility study, presented in \cite{DiZenobio} and used in the present study, has been focused on the H-mode reference scenario (BT = 7.5T, I$p$ = 6.5 MA, other engineering parameters available in \cite{Crisanti} or \cite{Ramogida}), which is one of the most challenging scenarios among those foreseen during FAST activities. It is worth noting that this scenario is characterized by a relatively short duration, having a flat top lasting for $\sim12$ s only. The time evolutions of the operative currents $I_{op}$ for the CS1 modules only (see later) are reported in Fig. \ref{fig:correnti}.

As it can be seen from Fig. \ref{fig:correnti}, the H-mode reference scenario can be divided in the following main phases: the charge of the coil, lasting 8 s; a fast ramp down of the field $B$ (with a very large $\dot{B}$ in the initial 0.04 s), lasting $\sim$8 s; the flat top, lasting $\sim$ 12 s; the final discharge of the coil, lasting other 8 s.\\
Based on the room available in the resistive design, the SC version of the CS coil has been conceived. It consists of a rectangular winding pack (WP), inserted into a stainless steel casing, wound from a rectangular conductor with an aspect ratio lower than 2, having long twist pitch values, low void fraction ($\sim 29\%$) and no central channel. These choices, mainly aimed at giving better load support to the Nb$_3$Sn strands and consequent limited performance degradation, derive by many measurement campaigns and studies performed in the last years \cite{degradation}. Note that the choice of Nb$_3$Sn for the SC strand is required by the maximum magnetic field (around 17.7 T, see Section \ref{sec:fields}) arising in the coil. The details of the design for all the 6 modules making the CS coil are reported in \cite{DiZenobio}; since the central modules (CS1U and CS1L) experience the more stringent conditions (highest field intensities), we limit our analysis to them. Their main characteristics, together with details of the WP, are reported in Table \ref{tab:datiCavo}.

\begin{table}%[tbp]
\small
\begin{tabular} {|l|c|}
\hline
\hline
Cable inner dimensions (mm$^2$) & 28 x 35.2 \\
\hline
Cable jacket thickness (mm) & 2.5 \\
\hline
Corner radius (mm) & 250 \\
\hline
Strand diam. (mm) & 0.81, \cite{Lu91} \\
\hline
$\lambda$=(1+Cu:nonCu ratio)$^{-1}$ & 0.5\\
\hline
Cabling pattern & 3x3x5x5x6 \\
\hline
\# strands (all SC) & 1350 \\
\hline
VF & about 29 \% \\
\hline
I$_{op}$ (kA) & 40.2 \\
\hline
B$_{peak}$ (T) & 17.7 \\
\hline
\hline
\end{tabular}
\caption{central modules CS (CS1U and CS1L) and WP main features \cite{DiZenobio}.}
\label{tab:datiCavo}
\end{table}

\subsection{Magnetic fields}
\label{sec:fields}

Another fundamental ingredient of the present analysis is the magnetic field intensity experienced by the SC cable. A complete 3-D electromagnetic analysis of the whole scenario, including all the field contributions, has been performed with the Opera-3D code. All the three magnetic field components have been computed in function of the z coordinate (the CS axis is chosen as z axis, with the origin in the equatorial plane separating the CS1L and CS1U modules) and of the radial coordinate (with full computations for inner, outer and middle radii of the CS module; in-between values has been linearly interpolated).
The time evolution has been described by linearly interpolating the values fully computed at selected instants of the transport current rate changes.
The maximum peak field 17.7 T is found at z=+12 cm from the equatorial plane. 
In the following we consider the pancake (belonging to the CS1U module) subjected to this peak field: being near the equatorial plane, the main contribution to the magnetic induction $B$ arises from the z-component, whereas the azimuthal one is zero and the radial one is small and hereafter neglected. 
Therefore, the whole pancake cable is assumed to be subjected to a unidirectional transversal field $B_z$. 
Each turn of the pancake has been considered to be subjected to a uniform field equal to the one actually present in its axis, i.e. evaluated at its median radius.

\subsection{Heat loads - AC losses}

The main heat load on the superconducting cable is originated by the applied time-varying electromagnetic fields, which cause power dissipation within the superconductor material as well as in the normal metal matrix and cable jacket. Neutron heating is neglected in CS coils since it is essentially fully screened by the straight legs of the TF coils surrounding the plasma chamber. 
It is well known \cite{Wilson} that dissipation in SC cables submitted to time-varying fields and transport currents consists of three main contributions: hysteresis losses, coupling losses, and the so-called "dynamic resistance" originated losses.

The hysteresis losses $Q_h$ occur in Type-II superconductors because of the irreversible magnetization, due to the pinning of vortices, which arises when the external magnetic field $B$ is swept. The main parameter governing these losses is the penetration field $B_p$ here defined as the field at which an applied increasing field fully penetrates the cable, starting from the virgin condition. The penetration field $B_p$ depends on the geometry, dimensions and critical current densities. It is also worth noting that $B_p$ depends on the temperature $T$ and on the external magnetic field through the critical current density.

The coupling losses $Q_c$ are due to coupling currents which arise in multifilamentary wires and multistage cables, flowing in the superconducting wires and closing through the normal metal (the matrix and the stabilization layer), where Joule dissipation actually occurs. These current are excited by the time-varying flux of the magnetic field threaded to the multifilamentary wire. The main parameter governing coupling losses is a characteristic coupling time constant $\tau$, which dictates the time scale of coupling current transients, and which depends on the strands/wires twist pitch and on the inter-filament/strand transverse resistivity. Actually, multistage cables introduce multiple $\tau_k$ (and corresponding volume fractions) for each stage of the twisting; here, as often done for simplicity sake, we consider a single effective time constant.

Finally, the ``dynamic resistance'' losses $Q_d$ occur whenever the current carrying cable is subjected to a time-varying field: the latter would change the transport current which has to be maintained by the current source. The work, and thus the dissipated energy, is then provided by the current source itself instead from the external magnetic field, as it happens for the two losses above discussed. This latter contribution is often incorporated in the hysteresis loss term, since they share a common nature.

In addition, one has to take into account losses due to eddy currents $Q_e$ in pure bulk normal metals: in our case, since the strands are all SC, the only normal metal is given by the cable jacket. 

According to \cite{Wilson} (and to \cite{ITER} regarding the eddy current losses), the expressions for the power dissipation per unit volume (W/m$^3$) in the static regime (slow field changes, see later) are:
\begin{equation}
\label{eq:Pc}
    P_c=n\tau\frac{|\dot{B_i}\dot{B_e}|}{\mu_0}
\end{equation}
\begin{equation}
\label{eq:Ph}
    P_h=\frac{2}{3}B_p (1-i^2) \frac{|\dot{B_e}|}{\mu_0}
\end{equation}
\begin{equation}
\label{eq:Pd}
    P_d=\frac{4}{3}B_p i^2 \frac{|\dot{B_e}|}{\mu_0}
\end{equation}
\begin{equation}
\label{eq:Pe}
    P_e=\frac{l_1^2}{12\rho_{jacket}}{\dot{B_i}^2}
\end{equation}
\noindent where the moduli ensure that the power losses are always positive, as they should be, for both increasing and decreasing fields; $i=I_t/I_c$ is the transport current normalized over the critical current, \footnote{The terms $(1-i^2)$ and $i^2$, present in $P_h$ and $P_d$ respectively, take into account the effect of the transport current. Note that Ref. \cite{Wilson} gives directly $(P_h+P_d)\propto(1+i^2)$}, $n$ is a geometrical factor (in general taken as equal to 2, more often incorporated in $n\tau$ as a single global quantity), $\rho_{jacket}$ and $l_1$ are the resistivity of the jacket and its outer dimension, normal to the field, respectively.
The quantities $B_i$ and $B_e=H/\mu_0$ are the induction field inside the cable and externally applied field evaluated inside the cable, respectively; the dot denotes the time derivative. 
In the static regime, the external and internal fields are related through the following expression \cite{Ries}:
\begin{equation}
\label{eq:BiBe}
    B_e-B_i=\tau\dot{B_i}
\end{equation}
The  expression for hysteresis losses assume that the actual external field swing is $\gg B_p$. 

As far as geometry is concerned, the system can be described in terms of (isolated) cylindrical strands in a transversal applied field: the corresponding penetration depth is $B_p=(1/\pi)\mu_0 J_c d_{eff}$, where $d_{eff}$ is the effective diameter of the superconducting filaments within the strand, and $J_c$ is the critical current density of the superconductor ($I_c=J_c A_{nonCu}=J_c \lambda A$, where $A$ and $A_{nonCu}=\lambda A$ are the total and nonCu strand sections, respectively). The numerical prefactors in Eq. \eqref{eq:Ph} and \eqref{eq:Pd} have been taken within the same geometry (isolated cylindrical strands in a transversal field).
For a line of cylindrical cables/strands close-packed along the field direction, the slab geometry (with the corresponding geometrical factors) could be a possible description \cite{Wilson}. Since the strands considered in the following have a significant thickness of the stabilizing matrix ($\approx 0.2$ mm), the filamentary zone within the individual SC strands (thickness $\approx 0.4$ mm) can be considered sufficiently separated ($\approx 0.4$ mm) in order to be satisfactorily approximated by the isolated cylinder geometry.

Finally, in order to obtain the power loss per unit length (W/m), $P_h$ and $P_d$ must be multiplied by $A_{nonCu}=\lambda A$, $P_c$ too (provided that $n\tau$ is correspondingly defined as to be referred to the $A_{nonCu}$ only), and $P_e$ by an effective jacket cross section $A_{jacket}=(l_1*l_2-l_3^3*l_4/l_1^2)$ \cite{ITER}.

As initially stated, the above model holds in the static regime, i.e. slowly varying fields with respect to the system characteristic time. With rapidly time-varying fields, the screening currents in the outer layers of the cable reach $J_c$; by further increasing $\dot{B_e}$, the thickness of the screening layer increases until all the superconducting filaments reach $J_c$. No further screening is then possible and the cable behaves like a solid monolithic conductor. A saturation parameter is introduced \cite{Ogasawara80} as $\beta=\tau\dot{B_i}/B_{p,strand}$, where $B_{p,strand}=(1/\pi)\mu_0 D_{eff}\lambda J_c$ is the penetration field of an ``equivalent'' strand, i.e. of a superconducting cylinder of diameter $D_{eff}$ having the same $J_c$ and magnetization of the cable. As already noted for $B_p$, it is worth noting that both $B_{p,strand}$ and $\beta$ depend on $B_i$ and $T$ through $J_c$.
The static regime occurs at $\beta\ll1$, whereas full saturation occurs at $\beta\gg1$. In the latter case, expressions for power losses are \cite{Ogasawara80, Oliva}:
\begin{equation}
\label{eq:Pcsat}
    P_c=\frac{4}{3\pi}B_{p,strand}(1-i^2)\frac{|\dot{B_e}|}{\mu_0}
\end{equation}
\begin{equation}
\label{eq:Phsat}
    P_h=0
\end{equation}
\begin{equation}
\label{eq:Pdsat}
    P_d=\frac{8}{3\pi}B_{p,strand} i^2\frac{|\dot{B_e}|}{\mu_0}
\end{equation}
\noindent where, in this limit, the following relation holds:
\begin{equation}
\label{eq:BiBesat}
    B_e-B_i=B_{p,strand}(1-i) {\rm sign}(\dot{B_e})
\end{equation}
For completeness, we mention that in the intermediate regime (given approximately by $\beta\lesssim1-i$), models are definitely cumbersome: only the model of Ref. \cite{Ogasawara80} allows to explicitly calculate all the quantities without complicated numerical computations, but on the other hand it fails to describe accurately the crossover between the two limiting regimes \cite{MuzziPhD}. 

\subsection{Cooling system}
The cooling path is taken in its simplest form: a double pancake cooling is considered (although, given the small length $\approx$ 42 m of a single pancake, also an hexapancake assembly could be devised). Therefore, the study can be focused on a single pancake (half of double pancake).
Typical values for the helium flow are considered: inlet temperature $4.4$ K, pressure 6 bar and helium flow 5 g/s.

\section{Simulation Results}
\label{sec:sim}

We compute the AC losses in the pancake located where the applied field $B_e$ is maximum. 
In order to simplify the computations, coherently with the scope of the present preliminary computations, we take $B_i\approx B_e=B$ (and therefore in $\dot{B_i}\approx \dot{B_e}=\dot{B}$). Since the considered $B_e(t)$ is piece-wise linear, in this way a small error is done, involving only small time intervals $\approx \tau$ after the instants where $\dot{B_e}$ changes. 
In the static regime, the main parameters of the model are the coupling time $n\tau$ (which governs the coupling losses) and the filament effective diameter $d_{eff}$ (which governs the hysteresis losses). We take $n\tau=55\;$ms (related to the non-Cu section of the strand), as experimentally determined in \cite{dellaCorte10} for the orientation of the cable with the long side parallel to the applied field, and $d_{eff}=50\;\mu$m as a value representative of the filament size in the SC strand here considered. 

The daring choice of all these parameters deserves some further comments: in order to meet the very demanding requirements of a large-size CICC (Cable In-Conduit Conductor) operating at 17.7 T and 40 kA, the characteristic properties of the most performing Nb$_3$Sn strands (non-Cu $J_c$(12 T, 4.2 K, -0.46$\%$)$\sim$1500 A/mm$^2$ \cite{Lu91}) available on the market have been considered at this stage of the design, together with conductor design parameters that have in principle been optimized for TF magnets operating conditions. 
Very well conscious of the limiting aspects of the present choice for a CS magnet, the study presented here should provide a baseline for the definition of the strand requirements (maximum $d_{eff}$) and conductor design (maximum cable $n\tau$), that can be accepted in order to fulfil the present design requirements. The final, optimized choices will need to be made as a trade-off between minimizing the heat load due to AC losses, while maintaining in the same time a sufficiently current carrying capability together with the stability of the performances against electro-magnetic cycling loads. So, for the time being the critical current density $J_c(B,T)$ is computed according to the formula given in Ref. \cite{Lu91}, taking a nominal strain $= -0.46\%$.
 
The saturation regime model, and the evaluation of the condition for its applicability, additionally require the value of the empirical parameter $D_{eff}$. This quantity, in multistage CICC, is difficult to determine, although it places itself in the range between the filament diameter $d_{eff}$ and the whole cable diameter \cite{Muzzi03}. In order to choose a reasonable value, we use as a first guess $D_{eff}\approx 0.81\;$mm, the diameter of the actual SC strand, following \cite{Muzzi03}.

The resistivity of the jacket is taken as $\rho=7.8\times10^{-7}\;\Omega$m, constant in the temperature range of interest, according to Ref. \cite{EFDA}. The losses due to the eddy currents in the jacket are easily evaluated and result - as it could be expected - negligible, so that in the following we will ignore them.

In order to set the temperature dependent quantities ($J_c$ for the AC losses, the densities and thermal capacities for the cooling part of the system), we take a constant and uniform $T=4.6\;K$. Obviously this is a crude approximation, but coherent with the scope of the this work, which is aimed to a preliminary assessment of the heat load in the CS superconducting cable. Refined simulations with thermo-hydraulic simulation codes such as Gandalf, taking into account the full time evolution and space distribution of the various physical quantities, will be performed in future works.

\begin{figure}[hbt!]
% Requires \usepackage{graphicx}
\centerline{\includegraphics[width=10cm]{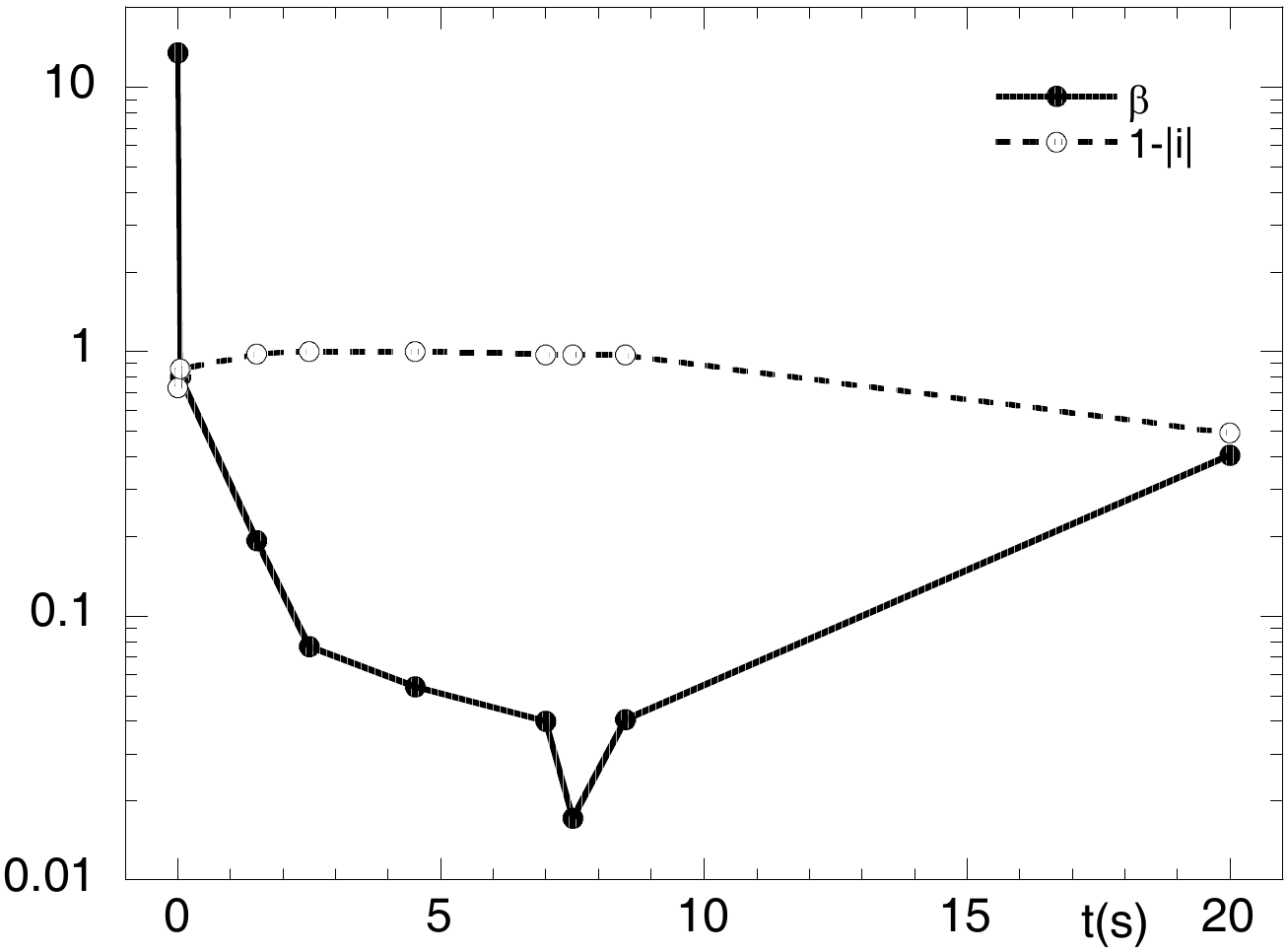}}
  \caption{Check of the saturation condition for the innermost turn: at t=0$\div$0.04 s the saturated regime is expected. (Lines are guides for the eye)}
\label{fig:saturazione}
\end{figure}

The choice of the model for the AC losses computation (static, saturated regime, or crossover) should be done according to the saturation parameter value. A straightforward evaluation of the saturation condition ($\beta>1-|i|$) shows that the innermost turns go into the saturation regime at the beginning of the field ramp-down (see Fig. \ref{fig:saturazione} for the innermost turn).

With the aim of yielding a preliminary evaluation of the AC losses, we choose to keep the used model as simple as possible, also in order to minimize the impact of a certain degree of arbitrariness of the approximations implicit in complex models.
The easiest solution would be to use the static model only, which would allow to have a worst case evaluation since, by neglecting the saturation occurrence, it overestimates the actual losses \cite{Ogasawara80, Muzzi03}.
Nevertheless, deep in the saturation regime the overestimation can become very large: for example, with Eq. \eqref{eq:Pc} a $P_c\approx40$ kW at $t=0\;$s on the first turn is easily obtained, instead of the $\approx3\;$kW given by the saturated limit expression \eqref{eq:Pcsat} and \eqref{eq:Pdsat}. In order to obtain more credible figures, we opt for the following choice. We use a piece-wise mixed model: each time that the local total power loss, evaluated in the static regime, is larger than the corresponding quantity within the saturation regime, the actual losses are taken as those given by the full saturation expressions. Obviously in this way the crossover region of pre-saturation remains still overestimated, but large overestimations are neverthless avoided.

\begin{figure}[hbt!]
% Requires \usepackage{graphicx}
\centerline{\includegraphics[width=10cm]{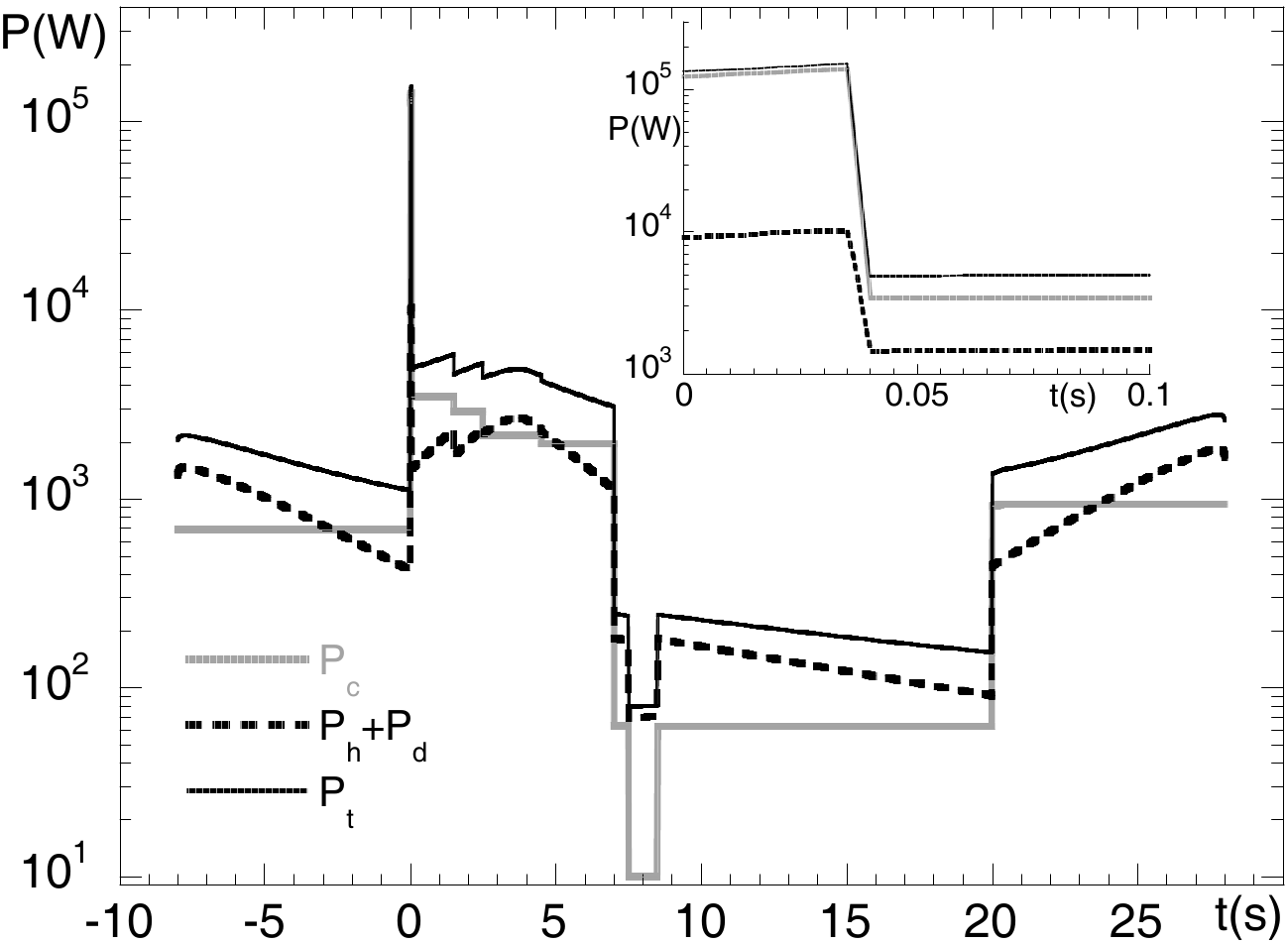}}
  \caption{Total power loss, and contributions due to various mechanims, against time in the pancake subjected to the largest field.}
\label{fig:powerloss}
\end{figure}

The plot of the total power losses in the selected against time is reported in Fig. \ref{fig:powerloss}.

It can be seen that the charge and discharge phases contribute with a significant heat generation, equal to 12.8 kJ and 16.1 kJ\footnote{Despite the specular symmetry of the $I_t(t)$ of the transport current in these two phases, the $B$ are slightly different, hence the difference in the energy dissipation.}, which is comparable with the one generated during the main phase.
Since the latter is by itself large and challenging for the coil (see later), the charge/discharge phases (and the corresponding heat loads) will no further commented, assuming that some sort of softening (i.e. a reduction of the field sweep rate and an eventual additional phase with a steady field - which would allow the SC cable to recover from the previously generated heating\footnote{in contrast with the original resistive coil, which dissipates also with constant fields.}) would be possible in the definition of the scenario.

We now focus on the main phase: the total energy dissipated through all the mechanisms in the whole $0-20\;$s time interval is about 39.3 kJ, which is definitely a large value with respect to the cooling power presumably available. The latter can be roughly evaluated by computing the $(mC)_{tot}=\sum_i m_i C_{i}$ of all the components of the cable (the Nb$_3$Sn and Cu fractions of cable, the steel of the jacket and the helium flowing in the void fraction of the cable), where $m_i$ and $C_i$ are the total mass and thermal capacity (at constant pressure) of  the $i^{th}$ component. Obviously the dominant contribution arises from the flowing helium, the physical properties of which (mass density and thermal capacity) are evaluated at $4.6\;$K, as already stated, and pressure 6 bar. The helium total mass is computed as the mass already present in the cable at $t=0\;$s plus the additional contribution arising from a standard flow of 5 g/s lasting for 20 s (the duration of the main phase). One obtains $(mC)_{tot}$=6.7 kJ/K, definitely insufficient to remove all the heat generated by the AC losses. Obviously this direct computation does not take into account the actual time evolution and space distribution of the heat load and cooling power, which make the scenario even more challenging. Such a full analysis, which has to be made by thermo-hydraulic simulation codes - as e.g. Gandalf, is out of the scope of this preliminary work which, as already stated, aims to preliminarily estimate the heat loads and possibly indicate the general directions of the design choices and changes needed to satisfy the project requirements.

\begin{figure}[hbt!]
% Requires \usepackage{graphicx}
\centerline{\includegraphics[width=10cm]{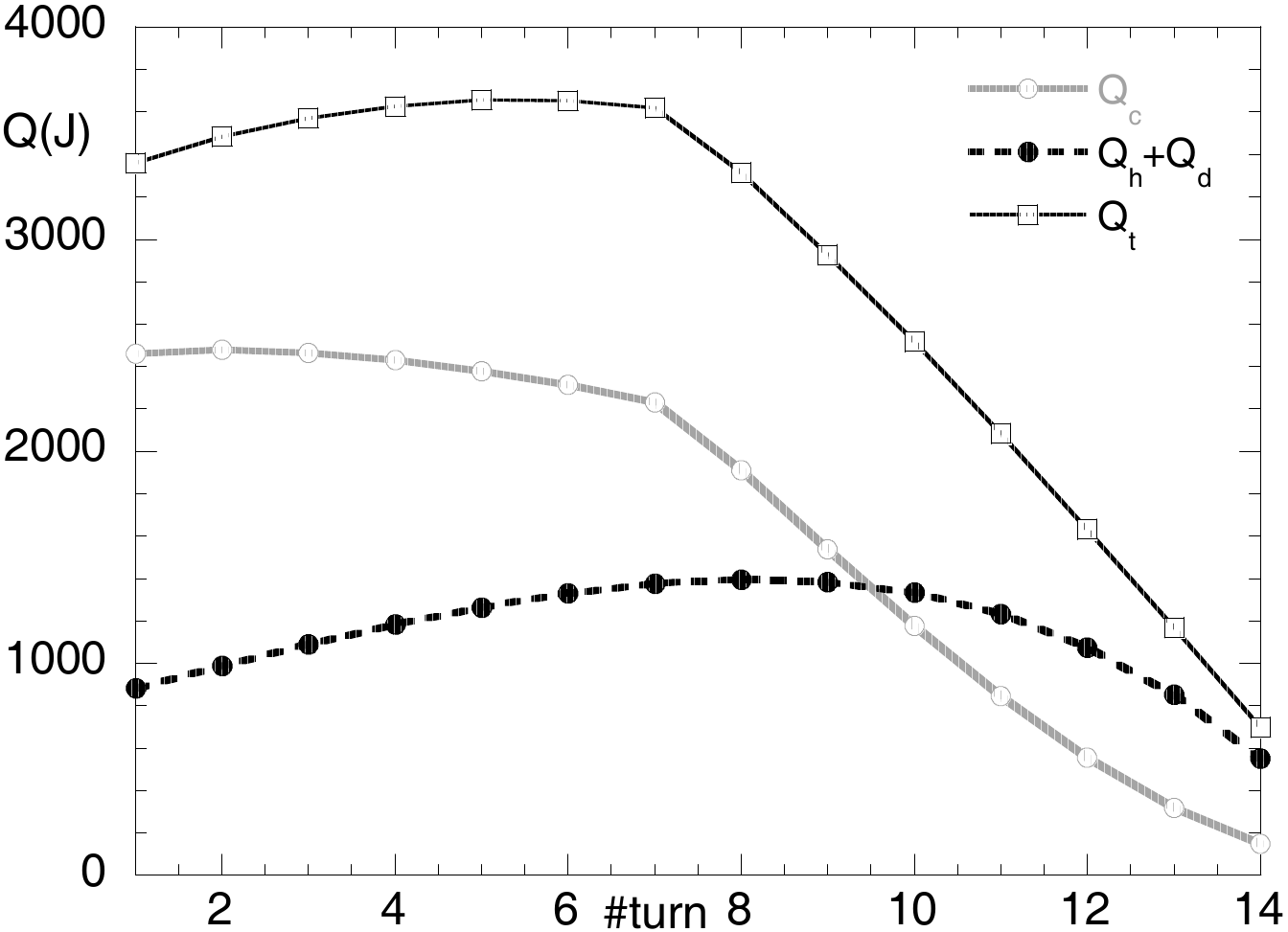}}
  \caption{Total dissipated energy, due to the various mechanism, for each turn.}
\label{fig:energyTurn}
\end{figure}

In this perspective, it is worth noting that total energy $Q_t$=39.3 kJ is the sum of $Q_h+Q_d$=15.9 kJ and $Q_c$=23.2 kJ (see also the energy dissipation distribution per turn reported in Fig. \ref{fig:energyTurn}).

This almost balanced splitting between coupling and hysteresis losses shows that, in order to significantly reduce the AC losses of the SC cable, both $d_{eff}$ (impacting on hysteresis losses) and $n\tau$ (impacting on coupling losses) have to be reduced.
For example, by assuming an ideal extremely high-performance cable, with $n\tau\sim10\;$ms and $d_{eff}=10\;\mu$m, one could reduce the total losses by a factor of 5, lowering them to 7.5 kJ.
By taking advantage of the shortness of the coil length, one could choose a higher helium flow (8 g/s), which would yield 10 kJ/K, so that the ratio $Q_t/(mC)_{tot}=7.5/10=0.75$ K is nearer to the needed operation conditions, but still unsatisfying. Indeed, the current sharing temperature $T_{cs}$ in the worst case, i.e. in the innermost turn at $t$=20 (so that one has the worst-case peak field 17.7 T and the maximum $I_t$), is $\sim 5.3$ K; if the above value of $0.75$ K gives a uniform heating of the SC cable, one would have 4.6 K + 0.75 K = 5.35 K $\gtrsim T_{cs}$, giving no temperature margin.

Going back to the instantaneous quantities, in Fig. \ref{fig:powerloss} it can be seen that coupling losses are dominant during the initial fast ramp down, whereas hysteresis losses are dominant in the flat top phase. The latter phase is the less challenging, although the dissipated power is still quite high, being on average $\approx$194 W, whereas the cooling power is around 19 W/K. A possible improvement of the heat balance can be obtained with the same criteria as before: by taking $d_{eff}=10\;\mu$m (main intervention) and $n\tau=10\;$ms, losses drop to 40 W, whereas the improved cooling gives 31 W/K. In order to sustain the flat top phase, therefore, large modifications of the cable should be taken into account; perhaps, also an increase of the cable void fraction to increase cooling should be considered (compatibly with other design requirements).

\begin{figure}[hbt!]
% Requires \usepackage{graphicx}
\centerline{\includegraphics[width=10cm]{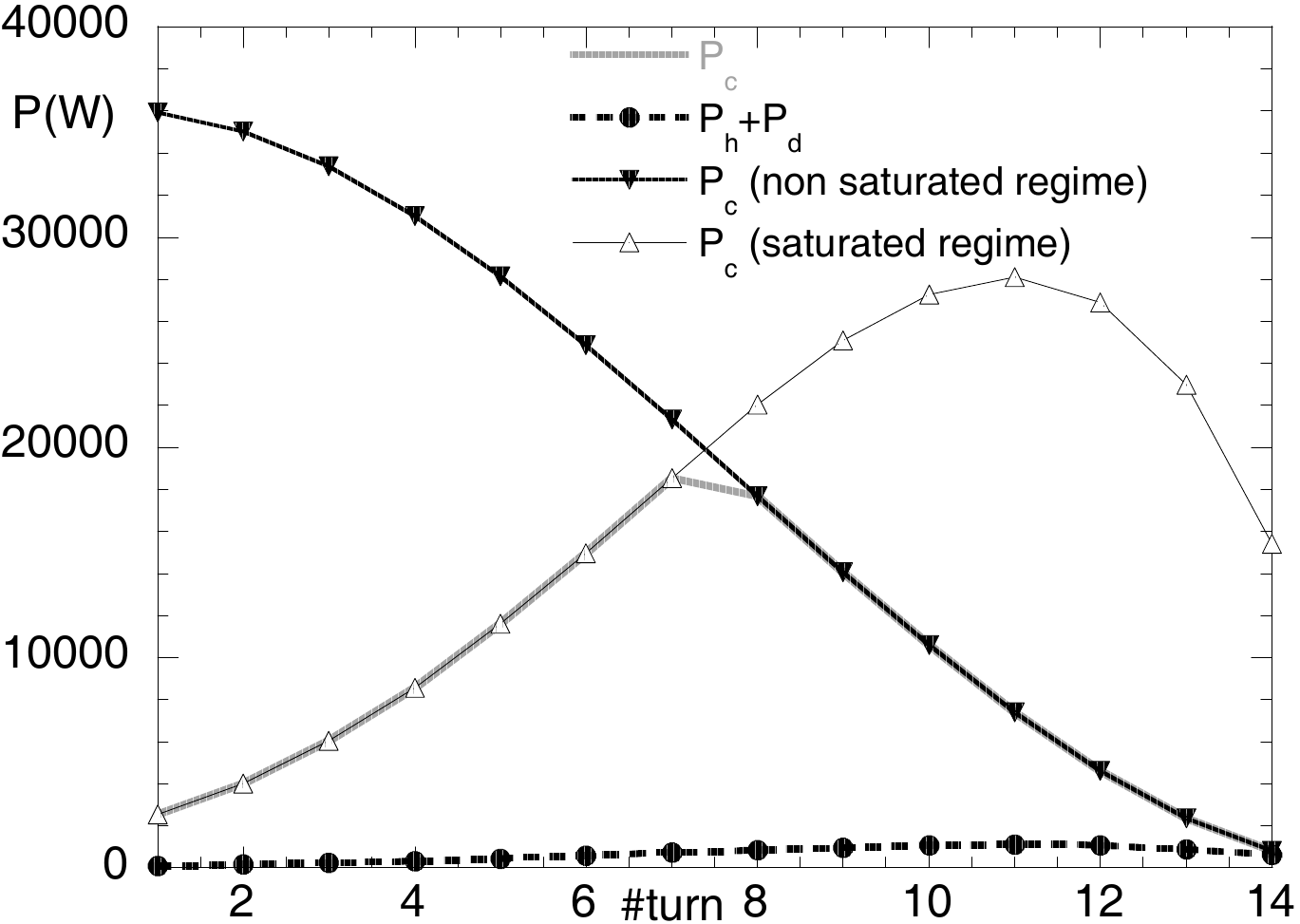}}
  \caption{Power loss at t=0 s per turn.}
\label{fig:initialPowerLoss}
\end{figure}

An even worst situation involves the initial ramp-down phase: an average 4.3 kW dissipation occurs, with a peak of 140 kW in the initial 0.04 s, caused by the very high $\dot{B}$ involved. A plot showing the distribution of the various kinds of power losses over the 14 turns of the pancake separately is shown in Fig. \ref{fig:initialPowerLoss}.

It can be seen that coupling losses are by far dominant. For comparison, the coupling losses computed within the static model only and the saturated regime model only are also reported. The wide maximum on $P_c$ around the 7$^{th}$ turn is due to the switch from saturated regime to the non saturated one: the losses in the nearby turns are definitely overestimated because of the rough modelization here used.

The roughness of the model obviously conspires against a sensible assessment of the actual margins: in these extreme conditions, in order to obtain trustworthy figures capable to guide the design process, both a full (and validated) model for the pre-saturated and saturated regimes is needed, as well as a reliable estimation of $D_{eff}$, which is the main scale factor governing both the entering in the saturated regime and the scale factor of the corresponding losses. 
Nevertheless, again an approximate consideration can be done: the total dissipated energy in the first 0.04 s is 
5.6 kJ. Given the extremely short time interval considered, this quantity can be reasonably compared with the cooling power due to the helium already present inside the cable. Since the latter yields 4.7 kJ/K (in nominal conditions), the two figures would yield an increase of temperature $\sim$ 1 K. By assuming the same ideal high-performance cable as before, hence by reducing by the usual 5 factor $n\tau$, $d_{eff}$ and also $D_{eff}$, one obtains a smaller dissipated energy 140 kW$/5\times0.04\;$s=1 kJ yielding an equivalent, more sustainable temperature increase of $\sim0.2$ K.
On the other hand, if one remains adherent to the presently available cable performances, this initial field sweep needs definitely to be slowed down: since the mission of the project dictated a given time derivative of the total $B$ flux threading the plasma chamber, one could envisage a $\dot{B}$ reduction by simultaneously increase the radius of the CS. This solution, however, should be weighted against the already very high constraint due to the peak magnetic field on the conductor.

\section{Conclusions}
The AC losses heat load for the CS module for FAST, in the hypothesis of a superconducting cable based design, has been evaluated within the very challenging H-mode reference scenario.
This preliminary analysis shows that the high fields and high field rates regimes drive the SC cable to high dissipations conditions, both near and well within the saturated regimes. Therefore, it is mandatory to have full, reliable models, together with a proper characterization of candidate SC cables at saturation, in order to properly address these regimes and correctly estimate the corresponding expected heat load. Only in this way a proper evaluation of the real cable requirements can be done.

Nevertheless, the present preliminary analysis allows to say that, in order to meet the stringent design requests, one has to consider much more performing strand (lower $d_{eff}$, but maintaining high Jc) and cable (lower $n\tau$ and lower losses in the saturation limit).
Even if some optimization of the cable design can be envisaged, most probably the field ramp rate should be reduced. More quantitative considerations will be provided in future works, in which full thermohydraulic simulations, based on the inputs of the present analysis, will be performed.

In closing, we note that the difficulties of the present superconducting design are, at least partially, originated from the choice to rigidly adhere to the characteristics and volume constraints of the standard FAST design, which is based on resistive magnets. If it were possible to change completely the design in a superconducting cable perspective, maintaining in the same time the mission of the project, the requirements on the SC cable would probably become less stringent.

\end{document}